\documentclass{aa}  

\usepackage{graphicx}
\usepackage[colorlinks]{hyperref}
\usepackage{caption}
\usepackage{subcaption}
\hypersetup{citecolor={blue}}
\usepackage{txfonts}
\begin{document} 

   \title{Kinematic and extinction analysis of a potential spiral arm beyond the Galactic bar}

   \author{Simran Joharle
          \inst{1,2},
          Francisco Nogueras-Lara\inst{3,4},
          Karl Fiteni \inst{5,6}
          }

   \institute{$^{1 }$University of Heidelberg, Heidelberg, Germany\\
              $^{2 }$Max Planck Institute for Extraterrestrial Physics\\
              \email{joharle@mpe.mpg.de},\\
             $^{3 }$European Southern Observatory, Garching, Germany\\
             $^{4 }$Instituto de Astrof\'isica de Andaluc\'ia (CSIC), Glorieta de la Astronom\'ia s/n, E-18008 Granada, Spain
             \email{fnogueras@iaa.es},\\
             $^{5 }$Como Lake centre for AstroPhysics (CLAP), DiSAT, Universit{\`a} dell'Insubria, via Valleggio 11, 22100 Como, Italy\\ 
             $^{6}$Institute of Space Sciences \& Astronomy, University of Malta,Msida MSD 2080, Malta\\
             }

\abstract
  {Determining the structure of the Milky Way is essential for understanding its morphology, dynamics, and evolution. However, studying its innermost regions is challenging due to high extinction and crowding. The detection of a double red clump (RC; core-helium-burning stars) feature at very low Galactic latitudes suggests the presence of a spiral arm beyond the Galactic bar, providing new insights into the Galaxy’s structure along this complex line of sight.}
  {We aim to evaluate the presence of this spiral arm by analysing the proper motion and extinction distributions of the detected RC features.}
  {We constructed proper motion and extinction difference maps to investigate the kinematic and reddening properties of the RC populations. We also confirmed the kinematic difference we see in observational data with N-body simulations of a Milky Way-like galaxy.}
  {We find that the two RC features are kinematically distinct, with a relative proper motion difference of $-0.16\pm0.02$\,mas/yr in the component parallel to the Galactic plane. This difference can be explained by Galactic rotation if the two RC features are located at different distances along the line of sight, consistent with our simulation results. The extinction towards the secondary RC is also $\sim0.05$\,mag higher than that of the primary RC. Additionally, we estimate that the extinction difference between the RC features corresponds to only $\sim5$\,\% of the total extinction from Earth to the first RC, suggesting little interstellar material between the farthest edge of the Galactic bar and the kinematically distinct structure traced by the secondary RC. As a secondary result, we derived the extinction curve using $JK_s$ photometry, obtaining $A_J/A_{K_s}=3.34\pm0.07$, consistent with previous studies of the innermost Milky Way. We find no significant variation of the extinction curve across fields or along the line of sight, within the uncertainties. The results are compatible with the secondary clump stars belonging to the spiral arm, although we cannot exclude that the population belongs to the disc.}
  {}


   \authorrunning{Joharle et al}
   \titlerunning{Kinematic and extinction analysis of a potential spiral arm beyond the Galactic bar}
   \maketitle
%

\section{Introduction}
A comprehensive understanding of the Milky Way's structure is essential for placing it within the broader context of Galactic morphology, dynamics, and evolution. Over recent decades, considerable effort has been devoted to accurately characterising the primary parameters of the Milky Way's spiral arm structure, including its number, shape, and inter-arm separation (e.g. \cite{Bland-Hawthorn:2016aa}, and references therein). Recent studies utilising trigonometric parallaxes and proper motions of molecular masers associated with young, high-mass stars suggest that the Milky Way exhibits a four-arm spiral configuration with additional segments and spurs \citep[e.g.][]{Vallee:2014aa,Reid:2019aa,Shen:2020aa}. Despite these efforts, the overall structure remains unclear (e.g. \citealt{Momany:2006aa}; \citealt{Hou:2014aa}; \citealt{Khoperskov:2020aa}).

The challenge in discerning the Milky Way's structure stems primarily from the Sun's position within the Galactic plane. This low Galactic latitude perspective is heavily obscured by interstellar dust, hampering structural studies of the Galaxy \citep[e.g.][]{Chen:2019aa,Kammers:2022aa}. In particular, the line of sight towards the Galactic centre (GC) is among the most challenging, as extreme cumulative extinction severely limits the observations \citep[with an average value of $A_V \geq 30$\,mag, corresponding to $A_{Ks} \geq 2.5$\,mag; e.g.][]{Nishiyama:2008qa,Schodel:2010fk,Fritz:2011fk,Nogueras-Lara:2021wj}. As a result, studying the Milky Way's structure along this line of sight is particularly difficult, and analyses of its stellar populations are largely restricted to the near-infrared regime, which is less affected by extinction.

\begin{figure*}
\centering
\includegraphics[width=\textwidth]{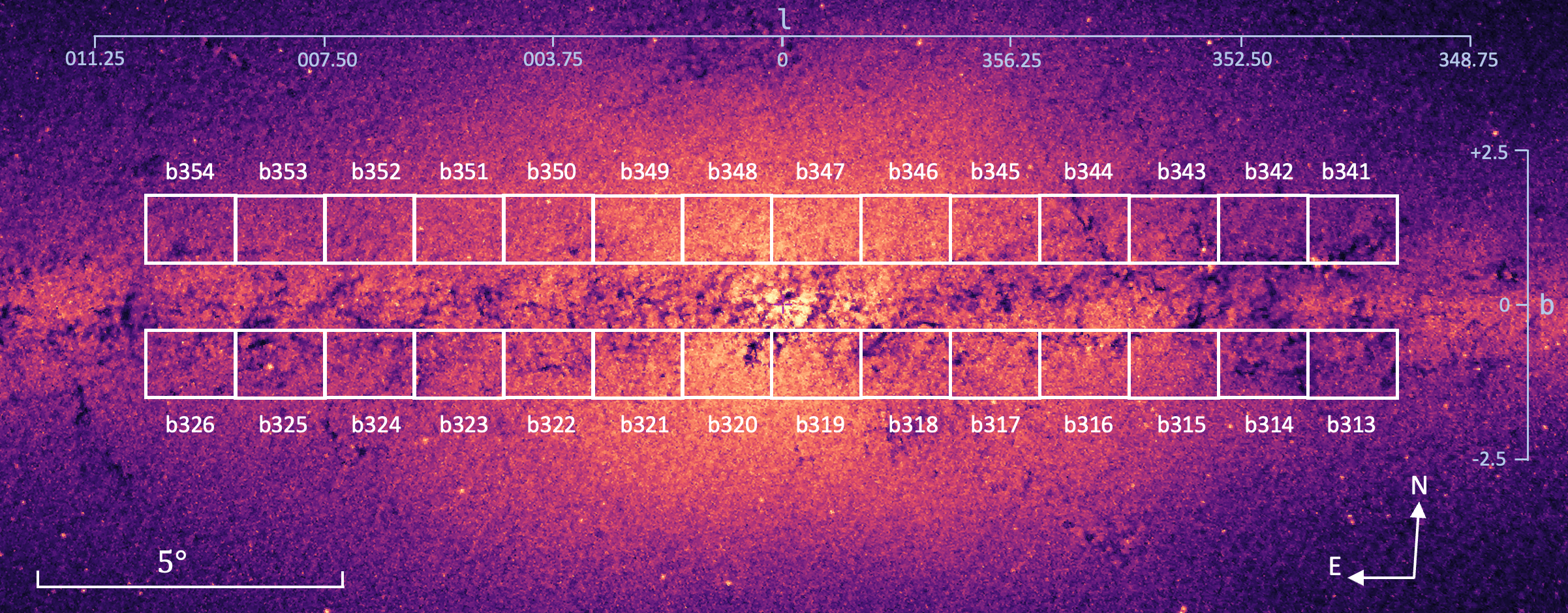}
\caption{2MASS $K_s$ image of the target region \citep{Skrutskie:2006te}, displaying the VVV survey tiles used in this study, covering $-10^\circ < l < 10^\circ$ and $-1.5^\circ < b < 1.5^\circ$, corresponding to the area analysed by \cite{Gonzalez:2018aa}. Each tile has dimensions of $1.208^\circ \times 1.487^\circ$. The scale is shown at the bottom left, and the directions for Galactic north and east are indicated at the bottom right.}
          \label{Tile_map}%
\end{figure*}

A powerful method for studying this complex line of sight is the analysis of red clump (RC) stars. These core-helium-burning stars have well-defined intrinsic properties \citep{Girardi:2016fk} and have only a modest variation in their intrinsic luminosity with age and metallicity, making them reliable distance indicators and valuable tracers of the inner structure of the Milky Way \citep[e.g.][]{Cabrera-Lavers:2008kx,Rezaei-Kh.:2018aa,Nogueras-Lara:2021uz, 2023A&A...673A..99U, 2022ApJ...931...72L}. In this way, \citet{Gonzalez:2018aa} constructed de-reddened luminosity functions around the RC feature, using photometry from the VISTA Variables in the Via Lactea (VVV) survey \citep{Minniti:2010fk}, allowing for a detailed study of the Galaxy’s innermost morphology. They identified a double RC in the luminosity function and attributed the secondary peak to an overdensity, likely linked to a spiral arm located beyond the Galactic bar. From their difference in magnitude, they estimated the secondary feature to lie at a distance of $\approx12-13$\, kpc from the Sun towards longitudes $l=0^\circ$, possibly coinciding with the location of the Perseus arm. This finding is further reinforced by RC density variations observed in the low-extinction VVV WIN 1733-3349 window, which align with the expected overdensities associated with a spiral arm beyond the Galactic bar/bulge \citep{Saito:2020vp}.

Nevertheless, previous studies have also linked similar secondary peaks to the presence of red giant branch bump (RGBB) stars, which belong to the same Galactic bar structure but represent a different evolutionary phase \citep[e.g.][]{Nataf:2011aa,Nataf:2013uq,Wegg:2013kx,Nogueras-Lara:2018ab}. The RGBB occurs when the hydrogen-burning shell interacts with a chemical discontinuity from the convective envelope’s first dredge-up, causing a temporary luminosity decrease and resulting in an accumulation of stars at a specific brightness \citep[e.g.][]{Khan:2018aa}.

This paper aims to understand whether the secondary RC stars detected by \cite{Gonzalez:2018aa} are located behind the Galactic bar/bulge, potentially tracing a spiral arm, thereby providing insights into the Milky Way's structure along this complex line of sight. We analysed the stellar kinematics and extinction properties of stars associated with both RC features. Our results show that the stars from each RC feature exhibit significantly different kinematics and extinction, supporting the hypothesis that the secondary RC originates from a spiral arm beyond the Galactic bar/bulge \citep{Gonzalez:2018aa}. The paper is organised as follows: Section \ref{sec2} describes the data used; Section \ref{sec3} outlines the method for distinguishing stars from the bright and the faint RC features (RC1 and RC2, respectively); Section \ref{sec4} presents a comparative analysis of their proper motions; Section \ref{sec5} examines extinction variations between the two features; and Section \ref{sec6} concludes the paper.

\begin{figure}
\centering
 \includegraphics[width=0.5\textwidth]{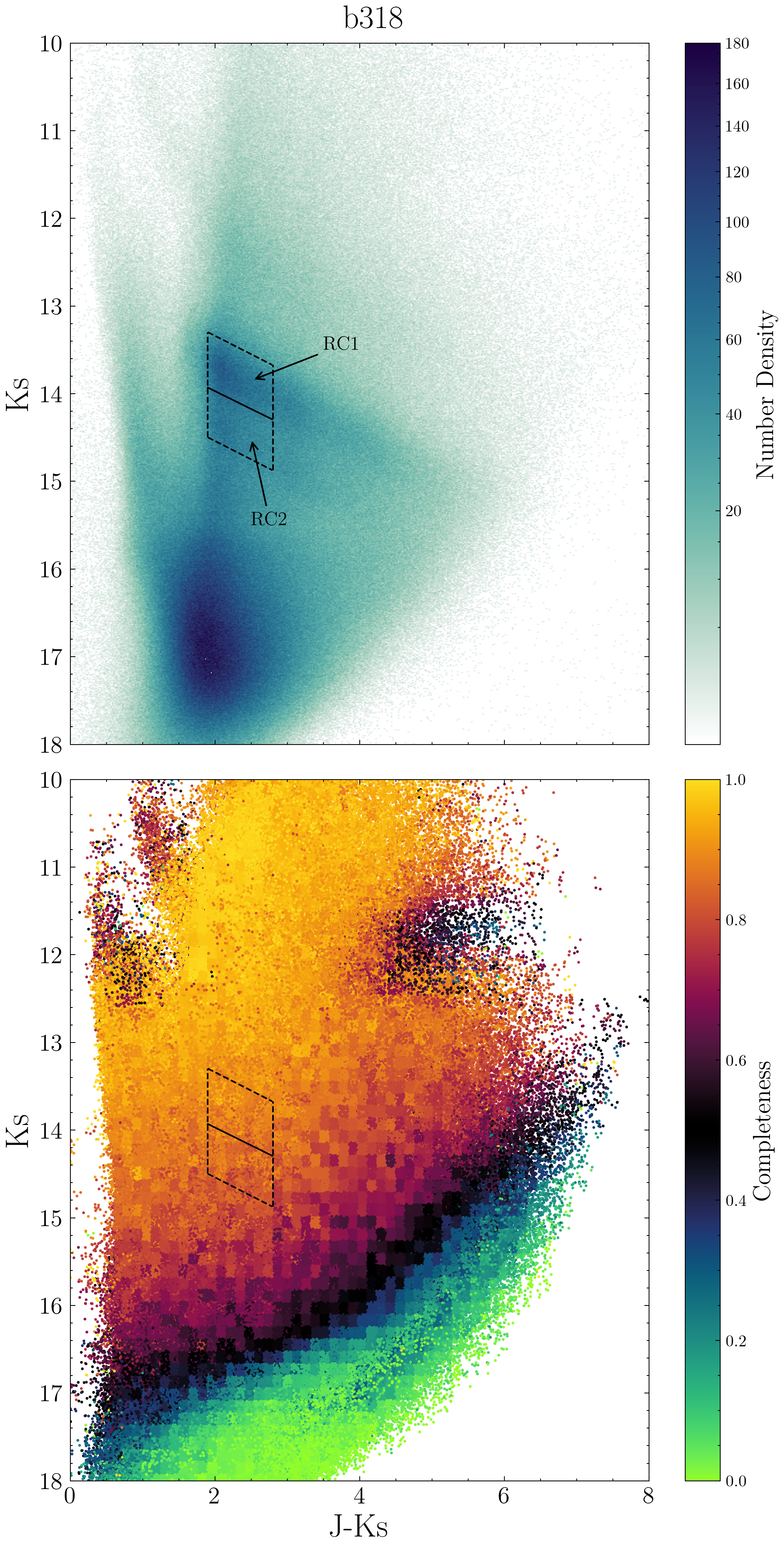}     
  \caption{Upper panel: CMD of $K_s$ vs $J-K_s$, with the selection box indicated, defined through colour cuts to exclude foreground stars and ensure completeness greater than 80\,\%. The line within the box represents the separation derived by fitting a line to stars exhibiting 50\,\% membership for each of the RC features (see Sect.\,\ref{sec3}). Lower panel: CMD, with each star colour-coded according to completeness values provided in \cite{Surot:2019aa}.
          }
     \label{CMD}
\end{figure}
\section{Data}\label{sec2}

We used data from the VVV survey, a near-infrared $ZYJHK_s$ photometric catalogue. The VVV survey covers 562 square degrees of the Galactic bar/bulge and an adjacent mid-plane region known for active star formation \citep{Minniti:2010fk}. Observations were carried out with the wide-field near-infrared camera VIRCAM \citep{Dalton:2006aa} on the 4 metre Visible and Infrared Survey Telescope for Astronomy \citep[VISTA,][]{Emerson:2010ab}, located at the ESO Paranal Observatory in Chile. Each VIRCAM observation generated a contiguous tile with a field of view (FOV) of $1.64\deg^2$.

For this study, we focused on tiles within the Galactic longitude range $-10^\circ < l < 10^\circ$ and Galactic latitude range $-1.5^\circ < b < 1.5^\circ$, as shown in Figure\,\ref{Tile_map}. These tiles match the region analysed by \cite{Gonzalez:2018aa}.

\subsection{Photometry}

We used the VVV point spread function (PSF) photometry available in the $J$ and $K_s$ bands from the deep $JK_s$ catalogue provided by \cite{Surot:2019aa}. This catalogue has limiting magnitudes of $K_s \approx 20$ and $J \approx 21$ and photometric completeness of RC stars ranging from 100\%-70\% across the Milky Way bar/bulge. For this study, we focused on the data from the 28 tiles shown in Figure \ref{Tile_map}, which correspond to the inner Galactic bulge region. 

The catalogue includes completeness information, which is crucial for our study due to the challenges of stellar crowding and high interstellar extinction. Completeness was assessed through artificial star experiments, where the completeness fraction represents the recovery rate of injected artificial stars within specific colour-magnitude bins. These experiments also provide estimates of total uncertainties, accounting for both systematic and photometric errors. Uncertainties are determined using a fourth-degree polynomial fit to the dispersion of input magnitude minus recovered magnitude per magnitude bin, as outlined in \cite{Surot:2019aa}.

\subsection{Proper motions}

For our proper motion analysis, we utilised version 1 (V1) of the VVV Infrared Astrometric Catalogue (VIRAC), developed by \cite{18:2018aa} and based on standard products from the v1.3 pipeline of the Cambridge Astronomical Survey Unit (CASU)\footnote{http://casu.ast.cam.ac.uk/vistasp/}. The VIRAC V1 catalogue provides relatively proper motions and parallaxes for stars observed in the individual pointings (pawprints) of the VVV survey. This astrometric catalogue comprises 119 million high-quality proper motion measurements, with 47 million of these having statistical uncertainties below 1\,mas/yr.

The creation of the VIRAC catalogue involves processing VVV data using aperture photometry. Telescope pointing coordinates are crossmatched within a 20 arcsecond radius, forming sequences of images, known as pawprint sets, for each field at different epochs. A total of 2,100 pawprint sets were generated, enabling independent proper motion determination. An iterative approach was employed to select a set of reference sources with proper motion values near the local mean. Proper motions for all stars were calculated relative to the average motion of these reference sources. The final proper motion values were obtained by applying inverse variance weighting to the individual pawprint measurements. This study uses proper motion data from the VIRAC catalogue for stars located in the region shown in Figure\,\ref{Tile_map}.

\begin{figure*}
\centering
\includegraphics[width=\textwidth]{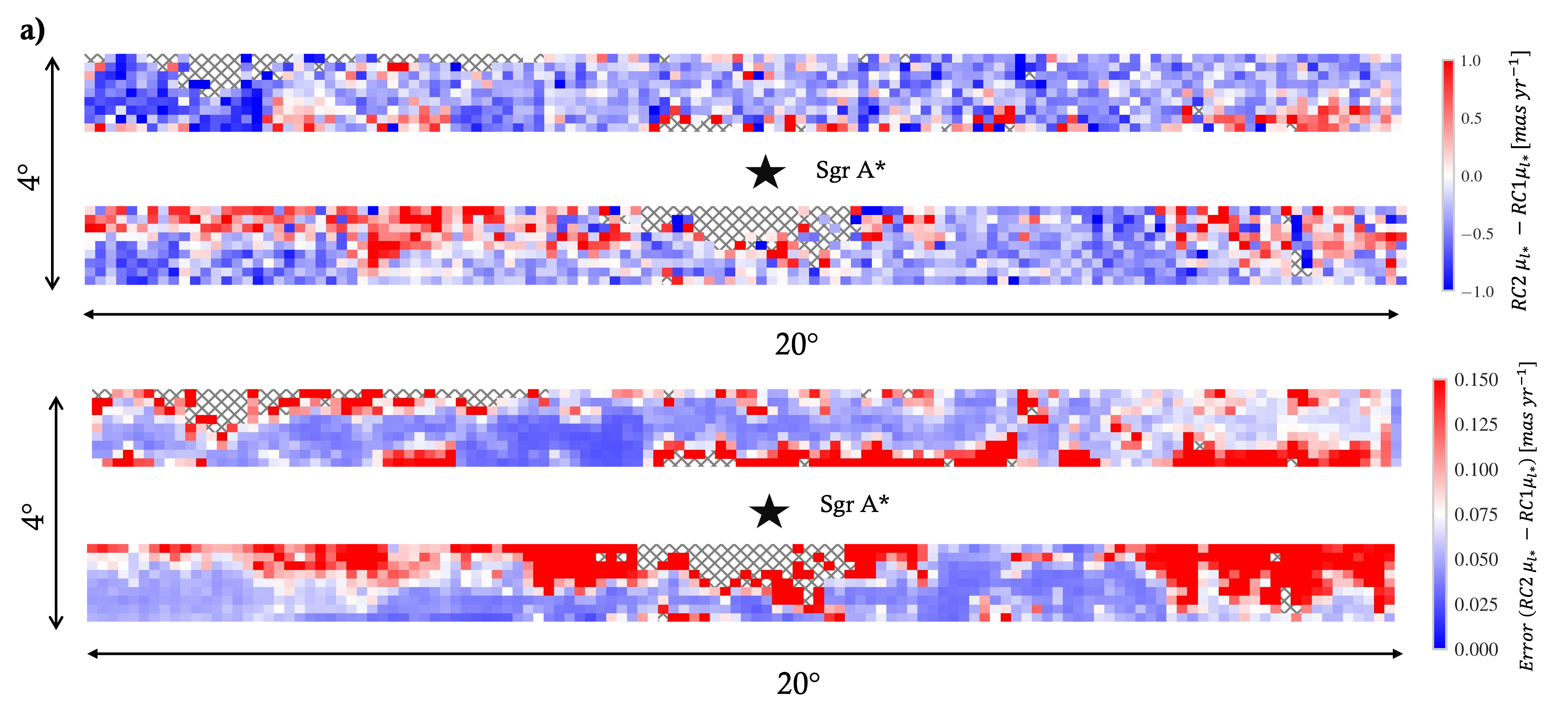}
\caption{$\mu_l*$ proper motion difference map (panel a) and associated uncertainty map (panel b), showing the difference between the mean $\mu_l*$ maps of RC2 and RC1. Colour bars indicate the proper motion values for each pixel (panel a) and their associated uncertainties (panel b). Empty pixels represent regions with an insufficient number of stars to compute a value. The position of the supermassive black hole, Sgr\,A$^*$, is marked by a black star in each panel, and the map sizes are indicated.}

\label{proper_map_mul}%
\end{figure*}

\begin{figure*}
\centering
\includegraphics[width=\textwidth]{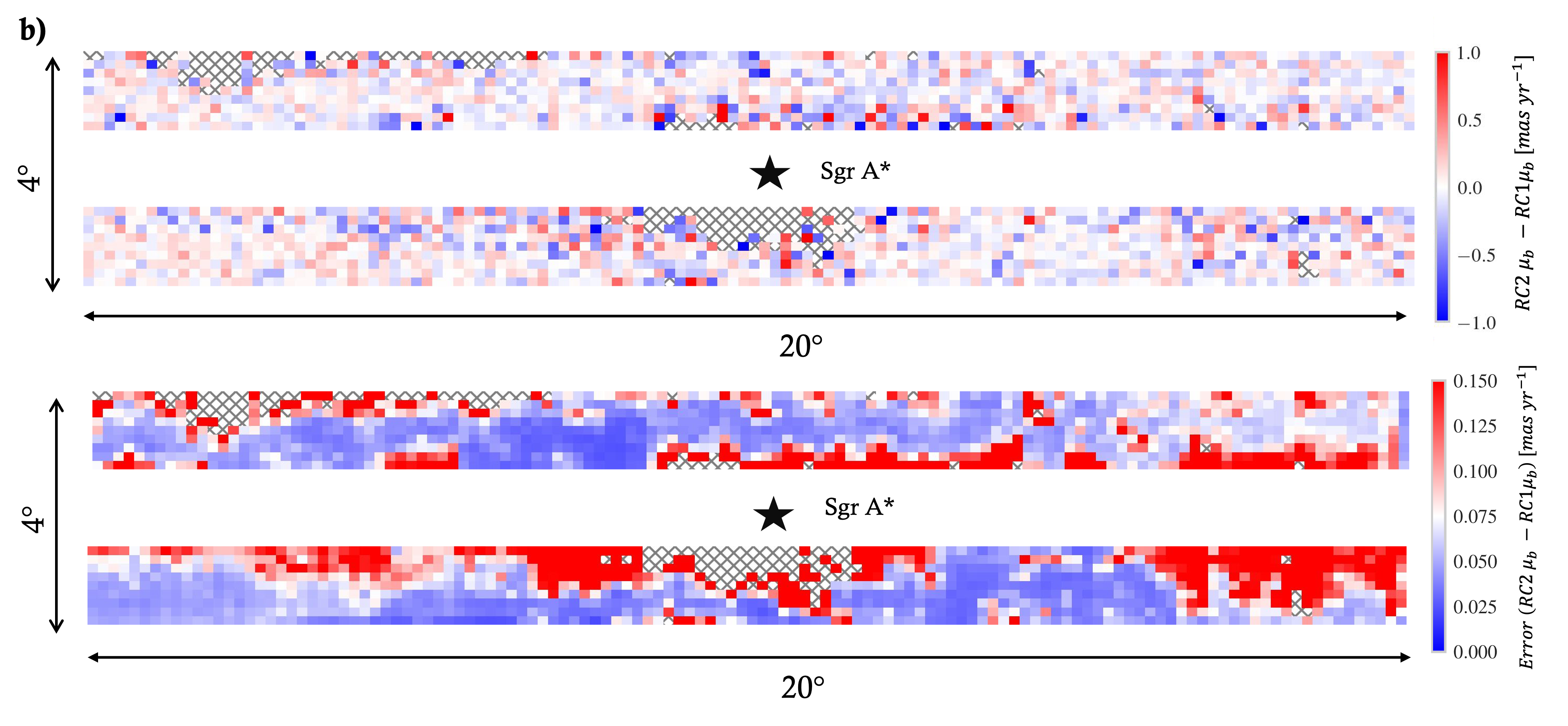}
\caption{$\mu_b$ proper motion difference map (panel a) and associated uncertainty map (panel b), showing the difference between the mean $\mu_b$ maps of RC2 and RC1. Colour bars indicate the proper motion values for each pixel (panel a) and their associated uncertainties (panel b). Empty pixels represent regions with an insufficient number of stars to compute a value. The position of the supermassive black hole, Sgr\,A$^*$, is marked by a black star in each panel, and the map sizes are indicated.}
          \label{proper_map_mub}%
\end{figure*}


\section{Colour-magnitude diagram and identification of RC features}\label{sec3}

To analyse the kinematics and extinction of the RC1 and RC2 stars, we first identified the stars belonging to these two features. We created a colour-magnitude diagram (CMD) and detected the features as overdensities in the RC region, consistent with the findings of \cite{Gonzalez:2018aa}. The top panel of Figure\,\ref{CMD} presents the $J-K_s$ versus $K_s$ diagram for one of the central tiles, b318, as an example.

We identified the stars belonging to the RC features for all the tiles by defining a selection box in the CMD. The upper and lower boundaries of the box are sloped at 0.42, in line with the reddening vector derived by \cite{Alonso-Garcia:2017aa}. The left boundary is set to exclude foreground stars, following the approach in \cite{Nogueras-Lara:2021wj}, by using the significantly different extinction of disc stars compared to those in the Galactic bulge. The colour cut used to isolate foreground stars is determined on a tile-by-tile basis, accounting for extinction variations for each line of sight. The right boundary is established to guarantee a stellar completeness greater than 80\%. The lower panel of Figure\,\ref{CMD} illustrates the completeness across different regions of the CMD and demonstrates how the left boundary is defined in the case of our example tile b318 to achieve this completeness threshold of 80\%.

We then identified the stars belonging to the RC1 and RC2 features within the defined selection box by subdividing it into 0.2\,mag colour bins. For each bin, we used the Gaussian mixture function from the SCIKIT-LEARN Python library \citep{Pedregosa:2011aa} to fit a 2D Gaussian model, matching the two features obtained by \cite{Gonzalez:2018aa} from the $K_s$ luminosity function. The separation between RC1 and RC2 is determined by the line that denotes the 50\% membership probability from the Gaussian mixture model (GMM) to belong to each of the Gaussian features. Stars above this separation line, which appear at brighter magnitudes in the CMD, are classified as RC1, while those below the line, at fainter magnitudes, are classified as RC2.

\section{Proper motion analysis}\label{sec4}

\subsection{Proper motions and photometry}

We analysed the proper motion distribution of RC1 and RC2 to assess whether these stars have different kinematics and then belong to distinct Galactic structures. To obtain the proper motion measurements, we crossmatched the selected RC1 and RC2 stars from the \cite{Surot:2019aa} catalogue with the VIRAC catalogue for each tile using TOPCAT \citep{Taylor:2005aa}. The crossmatch was performed with a 1.0 arcsec matching radius, selecting the closest match to ensure a one-to-one correspondence. Stars with proper motion uncertainties exceeding 1\,mas/yr were excluded. To avoid duplicates in overlapping regions, stars were selected from a single tile in these regions. The final proper motion catalogue contains approximately 30\,\% of the stars in the photometric catalogue, which have a well-defined proper motion counterpart.

\subsection{Proper motion difference maps}

To compare the proper motions of the stars in the RC features, we built proper motion difference maps for both components: one parallel ($\mu_l* = \mu_l \cos b$) and one perpendicular ($\mu_b$) to the Galactic plane. We first created separate proper motion maps for RC1 and RC2 stars, adopting a pixel size of $8.8' \times 7.2'$, which provided approximately 80 pixels per tile. A minimum of approximately ten stars per pixel was required for each RC feature map to estimate a proper motion value. This threshold, set at the fifth percentile of the star count distribution per pixel, was chosen to ensure statistical robustness. Pixels with fewer stars than this threshold were assigned `Not a Number (NaN)' values. We then computed 3$\sigma$ mean values for pixels with a sufficient number of stars.

We subtracted the RC1 maps from the RC2 maps to generate difference maps for the studied area. Figures \ref{proper_map_mul} and \ref{proper_map_mub} display the resulting maps for $\mu_l*$ and $\mu_b$, respectively. We also created the corresponding uncertainty maps by quadratically propagating the uncertainty associated with each pixel in the RC1 and RC2 proper motion maps. In each pixel the uncertainty of the mean proper motion was estimated as the standard error of the mean, i.e. the standard deviation of the proper motion measurements within the pixel divided by the square root of the number of stars contributing to that pixel. This approach did not incorporate the individual proper motion uncertainties of the stars, as we focused on measuring the statistical properties of the distribution. To assess the impact of individual measurement errors, we repeated the analysis using error-weighted means for each pixel, where each star’s contribution to the mean was weighted by the inverse square of its proper motion uncertainty. We then re-calculated the difference maps and found that the results remained consistent, confirming the robustness of our findings to the adopted averaging method.

The $\mu_l*$ difference map (Fig.\,\ref{proper_map_mul}) shows a significant shift towards negative values, indicating a negative difference between the $\mu_l*$ values in RC1 and RC2. An accumulation of `NaN' pixels is observed in the central region of the map, which can be attributed to the presence of the central molecular zone \citep[a dense accumulation of gas and dust that produces a significantly higher extinction in the innermost regions of the Galaxy; e.g.][]{Nishiyama:2008qa, Gonzalez:2012aa,Nogueras-Lara:2021wj,Henshaw:2022vl}. Some positive values are detected in the $\mu_l*$ difference map, but these are located near regions affected by extreme extinction and have large associated uncertainties, making them unreliable.

We estimated a 3$\sigma$ clipped mean value for the $\mu_l*$ difference map, obtaining a difference of $-0.16\pm0.02$\,mas/yr between the RC1 and RC2 stars. The negative difference can be explained by Galactic rotation, under the assumption that RC1 stars belong predominantly to the Galactic bulge and that RC2 stars are, on average, located farther away. A similar effect would also be observed if RC1 were split into stars at different distances, as stars located farther along the line of sight naturally exhibit different proper motions due to Galactic rotation. This behaviour is well established in simulations (see next subsection) and has also been observed in data (e.g. Soto et al. 2023). In any case, the significant proper motion difference between RC1 and RC2 indicates that they are not located at the same distance. This rules out the possibility that RC2 stars can be explained solely as RGBB stars \citep{Nataf:2013uq} belonging to the same stellar population as RC1; this result is consistent with the conclusion of \cite{Gonzalez:2018aa}.

On the other hand, the difference map for $\mu_b$ appears nearly uniform, with most pixels displaying values close to zero. We also observe the presence of a cluster of `NaN' pixels close to the centre, which can be equally explained by the extreme extinction. We estimated a 3$\sigma$ clipped mean value for the $\mu_b$ difference map, obtaining a difference of $0.0042\pm0.002$\,mas/yr. Our results effectively suggest no difference between RC1 and RC2 stars in the proper motion component perpendicular to the Galactic plane.

We validated our results by conducting several tests: 1) varying the minimum star count required to assign a proper motion value to each pixel, 2) computing both mean and median values instead of using the 3$\sigma$ clipped mean, and 3) testing different pixel sizes. Our conclusions remain consistent across all of these variations.

\subsection{Simulations}
\label{simulations}
\begin{figure*}
\centering
\includegraphics[width=\textwidth]{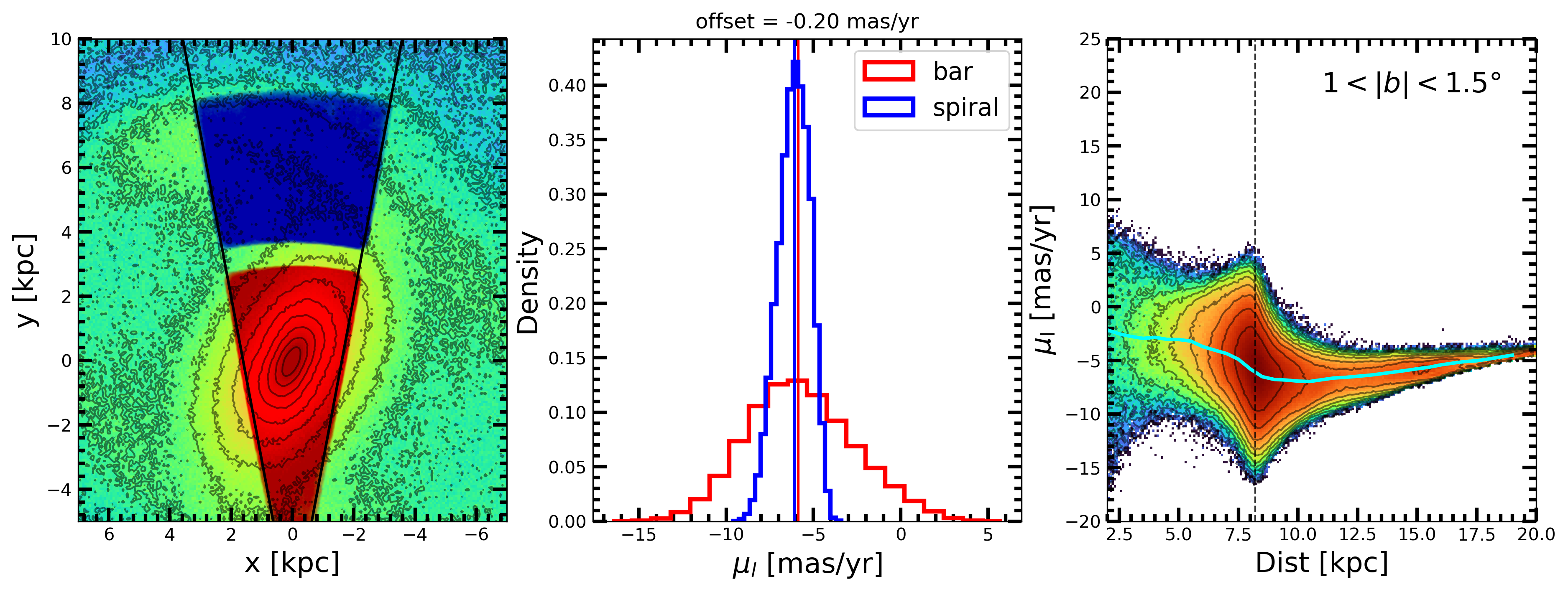}
\caption{Left Panel: Density distribution of stars used in the simulation analysis. The stars analogous to RC1 in red ($3 < d < 11$\,kpc) and those to RC2 are shown in blue ($12 < d < 16.5$\,kpc). They lie within $1\deg < |b| < 1.5\deg$. Middle Panel: $\mu_l*$ distribution for RC1 and RC2 with the median marked with solid lines. Right Panel: Stellar density distribution in $\mu_l*$ - d space. The dashed black line shows the centre of the model, while the cyan curve represents the median $\mu_l*$ in each distance bin.} 
          \label{Simulation results}%
\end{figure*}

To validate our findings, we used an N-body simulation of a Milky Way-like galaxy to examine whether a similar offset in $\mu_l*$ can be obtained among stars at different distances along the line of sight. We employed the model R1 from \cite{2005ApJ...628..678D}, which is a disc galaxy that forms a strong box/peanut-shaped (B/P) bulge through buckling instabilities of the bar. The model is scaled to the Milky Way as described in Gardner et al. 2014. We applied the performed transformations to obtain the proper motions, adopting a Solar position of $$(x,y,z) = (0,-8.2,0.012)$$ kpc and rotated the bar to an orientation of $27\deg$ relative to the Sun-GC line.
In the left panel of Fig.\,\ref{Simulation results}, we present the face-on stellar distribution of the model. The solid line reflects a FOV of $22\deg$. Within this FOV we categorised stars into two sub-samples: those associated with the bar (analogous to RC1), defined as stars with distances $3 < d < 11\,kpc$ marked in red; and those associated with the spiral arm (analogous to RC2), with distances $12 < d < 16.5\,kpc$, shown with blue markers. Consistent with our observational sample, we only considered stars with $1\deg < |b| < 1.5\deg$. The middle panel of Figure \ref{Simulation results} shows the $\mu_l$ distribution for these two populations, with the solid lines indicating the median $\mu_l$. For our selected samples, we retrieve a $\mu_l$ offset of -0.20\,mas/yr, comparable with the observational data. The right panel shows the stellar density distribution in $\mu_l$ - d space. The dashed black line shows the centre of the model, while the cyan curve reflects the median $\mu_l$ in each distance bin. On the far side of the disc, $\mu_l$ approaches zero with increasing distance. Therefore, any two stellar populations at different distances along the line of sight will have an offset in $\mu_l$. While we find that the sign and magnitude of the offset are sensitive to the spatial cut made for the two populations, the result shows that an offset in $\mu_l$ between two populations at different distances can naturally arise in a Milky Way-like potential, supporting our observational findings. 

\subsection{Profiles of proper motion differences}

\begin{figure}
\centering
\includegraphics[width=0.5\textwidth]{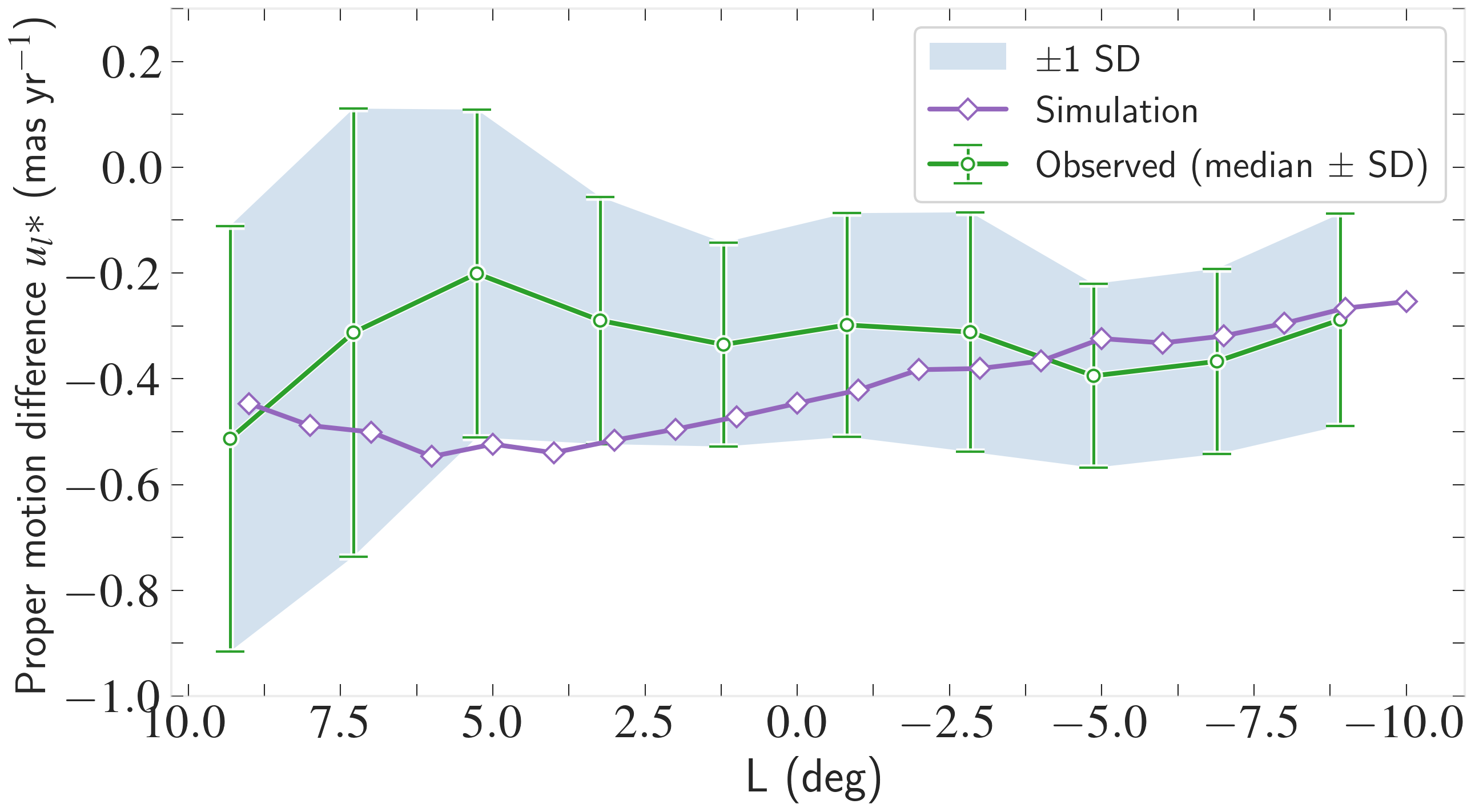}
\caption{Longitude profiles of the proper motion difference ($\mu_l^*$) between RC1 and RC2 stars. The green profile shows the observational data, obtained as described in the main text, while the purple profile shows an analogous synthetic profile from our simulation in Sect.\,\ref{simulations}.
}
          \label{mu_l profile test}%
\end{figure}

If RC2 stars belong to a spiral arm beyond the GC, the proper-motion difference between RC1 and RC2 should vary with longitude, reflecting the different rotation speeds of the Galactic components. To test this, we used the simulation described in Sect.\,\ref{simulations} to create a synthetic profile, excluding regions with $|b|<0.5^\circ$ to match the observational data. Figure\,\ref{mu_l profile test} shows the resulting profile together with the real measurements. The observational profile was obtained by binning in longitude and computing the median $\mu_l*$ difference in each bin. To minimise potential biases from regions with large $\mu_l*$ uncertainties—typically associated with high-extinction areas—we excluded pixels with $err(\mu_l*) > 0.075$\,mas\,yr$^{-1}$. 

The observational profile broadly follows the theoretical one, although the uncertainties remain large, mainly due to the limited number of stars in some tiles. Overall, this test supports the interpretation that the observed $\mu_l*$ differences arise from a more distant stellar population, potentially tracing a spiral arm along the line of sight. We also note that removing pixels with $err(\mu_l*) > 0.075$\,mas\,yr$^{-1}$ changes the sigma-clipped mean value of the map to $-0.308 \pm 0.0017$\,mas\,yr$^{-1}$.

\section{Extinction analysis}\label{sec5}

\begin{figure}
\centering
\includegraphics[width=0.5\textwidth]{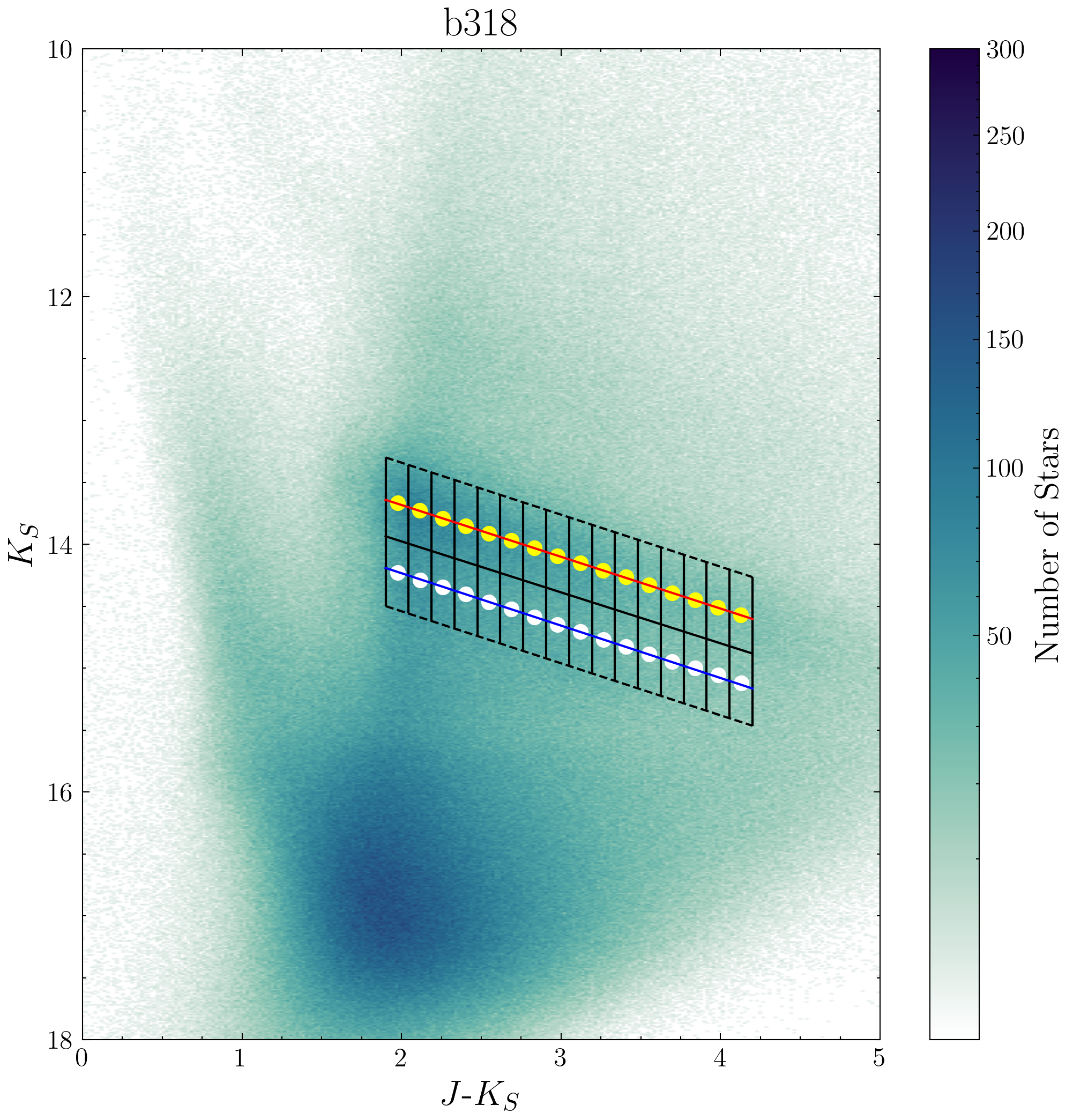}
\caption{CMD $K_s$ vs $J-K_s$ for the b118 tile, illustrating the slope determination for RC1 and RC2. The black parallelogram outlines the selection box for the RC stars, with divisions indicating the colour cuts used for the GMM analysis. Yellow and white points represent the mean values of the Gaussian models for RC1 and RC2, respectively. The red and blue lines show the fitted linear trends for each feature, from which the slopes are derived. The colour bar represents stellar density in the diagram.}
          \label{CMD for slope analysis}%
\end{figure}

\begin{figure}
\centering
\includegraphics[width=0.4\textwidth]{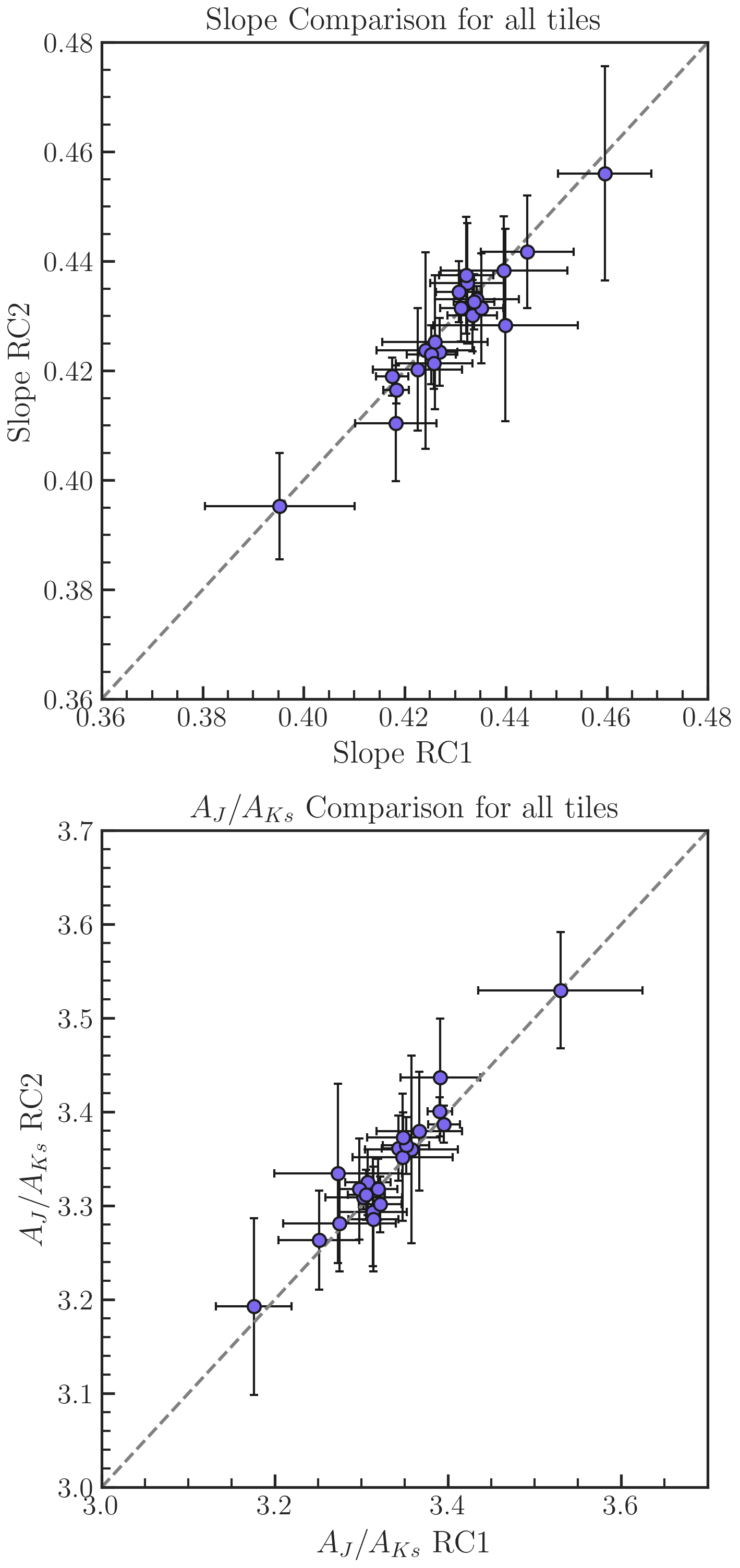}
\caption{Upper panel: RC2 slope vs RC1 slope for all the tiles with the associated uncertainties used in the extinction law calculations. A one-to-one line is plotted to indicate the good agreement between the values for each RC feature. Bottom panel: RC2 $A_J/A_{K_s}$ vs RC1 $A_J/A_{K_s}$ and the associated uncertainties. An analogous one-to-one line is also depicted.}

          \label{Extinction_Law}%
\end{figure}

\begin{figure*}
\centering

\includegraphics[width=\textwidth]{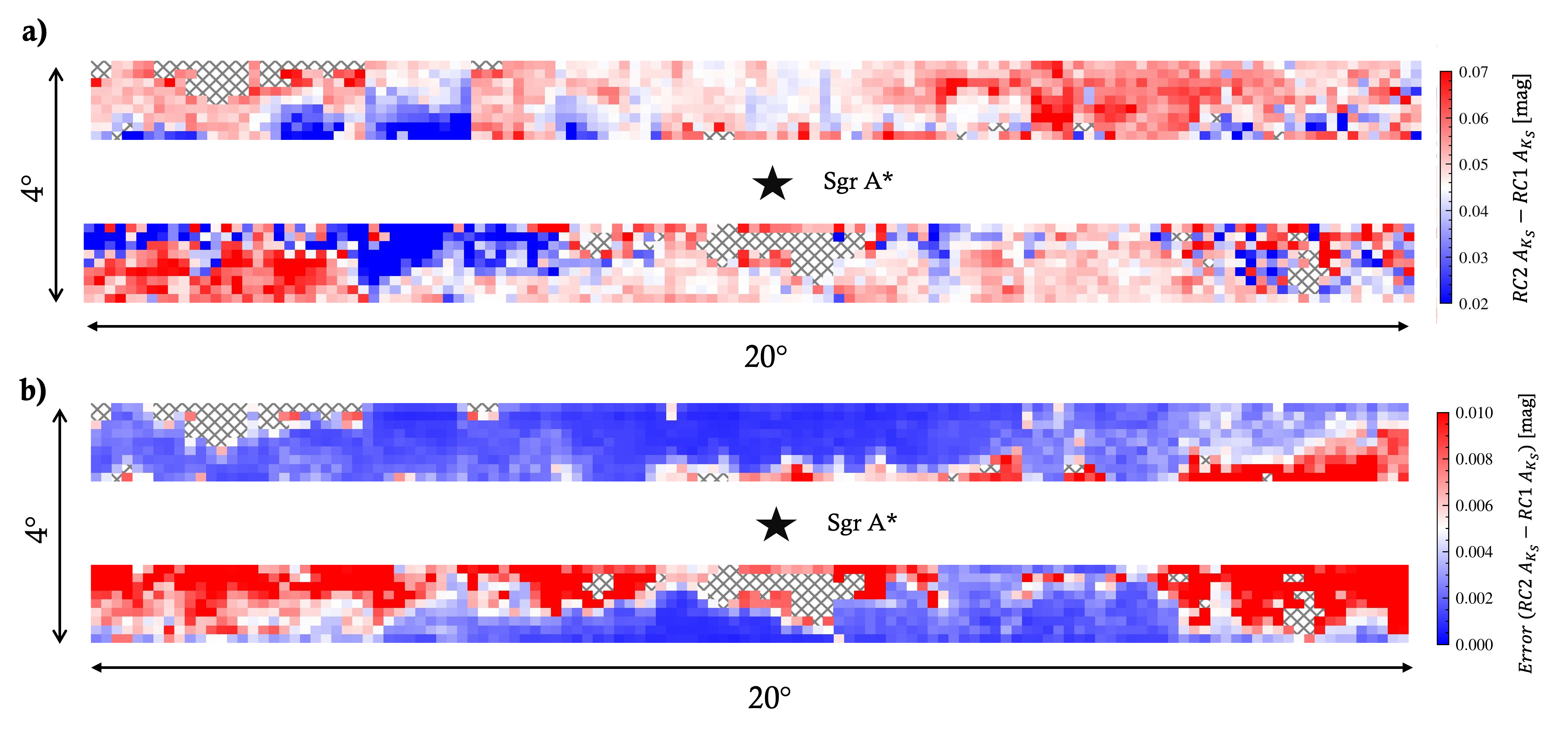}
\caption{Extinction difference map obtained by subtracting the extinction map of RC1 from that of RC2 (top panel), along with the corresponding uncertainty map (bottom panel). The colour bars indicate the extinction values for each pixel and their associated uncertainties. Empty pixels correspond to regions with an insufficient number of stars to compute a value. The position of the supermassive black hole, Sgr\,A$^*$, is marked with a black star in each panel, and the map sizes are indicated.}
          \label{Extinction_map}%
\end{figure*}

To complement our kinematic analysis, we examined the extinction towards each RC feature to characterise the reddening difference between the two stellar components with distinct kinematics. For this, we constructed an extinction difference map following a similar approach to that described in Sect.\,\ref{sec4}.

\subsection{Extinction curve}

Computing the reddening towards each RC feature requires adopting an extinction curve. However, the near-infrared extinction curve reported in the literature has changed significantly over the years, with the extinction index varying from values around 1.5 to beyond 2.0, or in some cases close to 2.5 \citep[e.g.][]{Rieke:1985fq,Draine:1989eq,Nishiyama:2006tx,Stead:2009uq,Gosling:2009kl,Fritz:2011fk,Alonso-Garcia:2017aa,Hosek:2018aa,Maiz-Apellaniz:2020aa,Sanders:2022wa,Butler:2024aa}. Recent studies also suggest that the near-infrared extinction curve does not follow a simple power law and instead varies with wavelength \citep[e.g.][]{Nogueras-Lara:2019ac,Nogueras-Lara:2020aa}. Additionally, the universality of the extinction law is not certain, given some reported variations depending on the location in the Galaxy \citep[e.g.][]{Zasowski:2009ys,Decleir:2022aa}. Given these uncertainties, we computed the extinction curve directly from the photometric data in our target region.

To derive the extinction curve, we used the RC1 and RC2 features, whose spread on the CMD is attributed to varying extinction, as they follow the reddening vector. We estimated the extinction curve as the ratio $A_J/A_{K_s}$ by combining Eqs.\,1 and 2 from \citet{Nogueras-Lara:2020aa}:

\begin{equation} 
\label{eqn2} 
A_j/A_{K_s} = 1 + \frac{1}{m}, 
\end{equation}

\noindent where $m$ is the slope of the reddening vector, determined from the slope of each of the RC features. To obtain these $m$ values, we used the method described in Sect.\,3.1 of \citet{Nogueras-Lara:2020aa}. Namely, we applied the previously outlined GMM approach to determine the mean values of the Gaussians corresponding to RC1 and RC2 in each 0.2\,mag $J-K_s$ colour bin. We then performed a linear fit to each RC feature and used a jackknife resampling method to estimate the slope. Figure\,\ref{CMD for slope analysis} illustrates this procedure with tile b318 as an example. Statistical uncertainties were derived from the Jackknife resampling method. To calculate the systematic uncertainties, we recomputed the slopes by considering different box dimensions and colour bin sizes. The total uncertainty was obtained by adding both in quadrature.

We excluded six tiles from this analysis (three at positive latitudes: b352, b353, and b354; and three at negative latitudes: b324, b325, and b326). In these tiles, the selection boxes on the CMD are narrower in colour than in the others, as the limits were set to ensure good completeness for both RC1 and RC2. The reduced colour range resulted in poorly constrained slope estimates with large uncertainties, hence giving unreliable extinction-curve estimates for these tiles. Table\,\ref{table_extinction} in the appendix shows the results for the slopes of the RC features and the extinction curve $A_J/A_{K_s}$ for all the tiles used for the extinction law calculations. 

Figure\,\ref{Extinction_Law} shows the comparison between the values obtained for each of the RC component in each tile. We found that the values agree well for both RC features in each tile and also across tiles. We computed the final value as the average of the values obtained for each RC and tile, obtaining $3.34 \pm 0.07$, where the uncertainty was estimated as the standard deviation of the distribution.

We did not detect significant variations in the extinction curve across the different tiles within the uncertainties, indicating that a single extinction curve can be assumed for the entire analysed region. This is consistent with previous findings for the GC \citep[e.g.][]{Nogueras-Lara:2019ac,Nogueras-Lara:2020aa}. Our extinction value also agrees, within the uncertainties, with previous estimates for the GC and the inner regions of the Galactic bulge (e.g. $A_J/A_{K_s} = 3.42\pm0.08$, \citealt{Alonso-Garcia:2017aa}; $A_J/A_{K_s} = 3.44\pm0.08$, \citealt{Nogueras-Lara:2020aa}; or $A_J/A_{K_s} = 3.23\pm0.11$, \citealt{Sanders:2022wa}). 

We also verified that the $A_J/A_{K_s}$ values obtained for each RC are similar. Using each RC feature independently to estimate the final value yields the same result: $A_J/A_{K_s} = 3.33\pm0.014$ for RC1, and $A_J/A_{K_s} = 3.34\pm0.017$ for RC2. This suggests that despite belonging to kinematically different stellar populations at different line-of-sight distances ($\sim$5\,kpc apart, according to \citealt{Gonzalez:2018aa}), the extinction curve does not vary significantly.

\subsection{Extinction difference map}

To construct the extinction difference maps, we first built separate reddening maps for stars belonging to the RC1 and RC2 features. The extinction ($A_i$) towards each star is calculated using Eq.\,A.1 from \cite{Nogueras-Lara:2021wj}:

\begin{equation} 
\label{eqn1} 
A_i = \frac{(j-i) - {(j-i)}_0}{(A_j/A_i)-1}, \end{equation}
\noindent where $j$ and $i$ are two Near Infrared (NIR) photometric bands ($J$ and $K_s$ in our case), the subindex $0$ represents the intrinsic colour, and $A_j/A_i$ is the extinction curve.

For the RC1 intrinsic colour, we assumed an old, metal-rich, and $\alpha$-enhanced population typical of the Galactic bulge \citep[e.g.][]{McWilliam:1994aa,Zoccali:2003aa,Lecureur:2007aa,Melendez:2008kx,Alves-Brito:2010aa,McWilliam:2010aa,Johnson:2014aa,Barbuy:2018aa}. Accordingly, we adopted $(J-K_s)_0 = 0.75$, as derived by \citet{Nogueras-Lara:2018ab} for an $\alpha$-enhanced stellar population of $\sim10$\,Gyr and twice the solar metallicity using BaSTI models \citep{Pietrinferni:2004aa,Pietrinferni:2006aa}.

In contrast, RC2 appears to trace a spiral arm, as suggested by \citet{Gonzalez:2018aa} and our kinematical analysis. The stellar population in this structure is expected to be less $\alpha$-enhanced \citep[e.g.][]{McWilliam:2016aa,Barbuy:2018aa}. Therefore, we adopted the intrinsic colour $(J-K_s)_0 = 0.62$, based on the general estimates for RC in the innermost regions of the Galactic bulge provided by \citet{Nogueras-Lara:2021wj}.

Following a similar approach to Sect.\,\ref{sec4}, we combined all tiles to construct extinction maps for RC1 and RC2 stars. Although the selection boxes for the RC features are different for each tile, the colour cuts are the same for RC1 and RC2 in each tile. Thus, the difference map, which is a relative map, is not affected in any way and shows the relative difference between both RC features regardless of the different selection boxes for each tile. Each map has a pixel size of $8.8' \times 7.2'$, with each pixel representing the mean extinction of the stars within it. To ensure statistical reliability, we set a minimum threshold for the number of stars per pixel. This threshold was based on the fifth percentile of the distribution of stars per pixel across the entire map, resulting in minimum counts of approximately $100$ stars for the RC1 and RC2 maps. Pixels with fewer stars were assigned `NaN' values. Additionally, we applied a 3$\sigma$ clipping algorithm to the stars in each pixel to remove potential outliers that could affect the extinction values. 

Figure\,\ref{Extinction_map} presents the resulting extinction difference map, computed as the extinction of RC2 minus that of RC1, along with the associated uncertainty map. The uncertainties were estimated by calculating the error of the mean for each pixel in the RC1 and RC2 extinction maps, then propagating them quadratically to obtain the corresponding values in the difference map. The extinction difference map is predominantly positive, indicating that RC2 stars generally experience higher extinction than RC1 stars. Empty pixels (`NaN') are concentrated near the GC, where the extinction is too high \citep[e.g.][]{Nishiyama:2006tx,Nogueras-Lara:2021wj} to observe stars beyond the Galactic bulge. This results in an insufficient number of stars to assign an extinction value, similar to what is seen in the proper motion difference maps in Sect.\,\ref{sec4}. A region with negative values was also identified, where RC1 appears more extinct than RC2. This region coincides with high-extinction areas in the reddening map from Figure\,1 of \citet{Gonzalez:2018aa}, suggesting that extreme extinction may lead to unreliable estimates. Furthermore, the larger uncertainties in this region (Fig.\,\ref{Extinction_map}, panel b) indicate that the results are not reliable.

Finally, we estimated the mean pixel value of the extinction difference map using a 3$\sigma$ clipping algorithm to discard unreliable pixels. This resulted in $\Delta A_{K_s} = 0.048\pm0.0001$\,mag, where the uncertainty was estimated by quadratically propagating the associated uncertainties of the pixels in the map.

To validate our results, we repeated the extinction difference map by: 1) varying the minimum star count required to assign an extinction value to each pixel, 2) computing both mean and median values instead of using the 3$\sigma$ clipped mean, and 3) testing different pixel sizes. Our conclusions remained consistent across all variations.

\subsection{Intrinsic colour and completeness effect}

In our extinction difference map calculation, we assumed that RC1 belongs to the inner Galactic bulge, while RC2 corresponds to a spiral arm with a stellar population characterised by lower $\alpha$ enhancement \citep[e.g.][]{McWilliam:2016aa,Barbuy:2018aa}. This implies different intrinsic colours, affecting the extinction estimates following Eq. \ref{eqn1}. Nevertheless, our results represent an upper limit on the extinction difference, since assuming a similar intrinsic colour for both stellar populations would lead to a smaller relative extinction difference.

Additionally, RC2 is significantly contaminated by RGBB stars from the Galactic bulge, the same stellar population traced by RC1. According to Table\,2 in \citet{Nogueras-Lara:2018aa}, these stars have an intrinsic colour of $(J-K_s)_0 = 0.70$\,mag, similar to that of RC1 stars. Their presence would imply a slightly redder intrinsic colour than the one we adopted for RC2, further reinforcing that the extinction difference we obtained represents an upper limit.

On the other hand, the RC features were selected to ensure high completeness (>80\,\%), even though RC2 is expected to have lower completeness than RC1, because it's a fainter feature. This could increase the relative extinction difference. To assess the impact of completeness, we also constructed extinction maps using only stars with available proper motions, for which completeness is significantly lower (only $\sim30\,\%$ of the stars in the photometric catalogue have proper motion measurements; see Sect.\,\ref{sec4}). This selection ensured that each RC feature was dominated by stars located at different distances, which exhibited distinct proper motions, so that the computed extinction difference would correspond to these two separate components. We obtained a mean extinction difference of $\sim$0.05\,mag, consistent with our main results.

\subsection{Extinction difference interpretation}

We interpret the positive extinction difference as evidence that RC2 stars lie farther from the Sun than RC1 stars. This is consistent with the observed proper motion offset and supports the view that RC2 traces a more distant stellar component—likely a spiral arm—beyond the Galactic bar, as suggested by \citet{Gonzalez:2018aa}.

Using the RC1 extinction map, we computed a mean extinction value of $A_{K_s} \sim 0.5$\,mag for RC1 stars, implying that the mean difference obtained from the extinction difference map in Fig.\,\ref{Extinction_map} accounts for only $5\,\%$ of this value. This suggests that, although RC2 experiences significantly higher extinction than RC1, overall extinction remains relatively low for a spiral arm located $\sim5$\,kpc beyond RC1 \citep[according to ][]{Gonzalez:2018aa}.

For comparison, \citet{Nogueras-Lara:2021uz} estimated the extinction towards the four spiral arms along the line of sight to the GC and found a total accumulated extinction of $A_{K_s}\sim1.2$\,mag for the 3\,kpc arm, the closest spiral arm to the GC. This arm is located $\sim5$\,kpc from Earth, comparable to the estimated distance between RC1 and RC2 \citep{Gonzalez:2018aa}, highlighting how small the extinction increase is between these two features.

However, we can justify the relatively low extinction difference we obtained, given that the fields analysed in this work are located $\sim2^\circ$ away from the GC (see Fig.\,\ref{Tile_map}), where the line of sight is less affected by extreme reddening \citep[e.g. see Fig.\,3 in][]{Gonzalez:2012aa}. Furthermore, the region between RC1 (associated with the Galactic bulge) and RC2 is expected to contain little gas and dust, as RC2 likely represents the first major structure encountered along the line of sight beyond the bulge, known to be low in gas and dust content.

\section{Conclusion}\label{sec6}

In this paper, we built upon the detection of a secondary RC feature in low-latitude VVV fields ($-10^\circ < l < 10^\circ$ and $-1.5^\circ < b < 1.5^\circ$, excluding the region within $b\in[-0.5,0.5]$), which was previously suggested to trace a spiral arm beyond the Galactic bar \citep{Gonzalez:2018aa}. We identified the stars belonging to each RC feature using CMDs and characterised their kinematics and extinction through proper motion and extinction difference maps. Our analysis revealed that RC2 stars exhibit a significantly different $\mu_l*$ distribution ($\Delta\mu_l*=-0.16\pm0.02$\,mas/yr) and mean extinction ($\Delta A_{K_s} = 0.048\pm0.0001$\,mag) compared to RC1 stars.

Our findings support the interpretation that RC2 is located beyond the Galactic bar, likely in a spiral arm. The significant differences in both kinematics and extinction between the two RC populations confirm that they belong to distinct structures, with RC1 associated with the Galactic bar/bulge and RC2 forming part of a more distant stellar component. This is consistent with the distance-modulus analysis carried out by \citet{Gonzalez:2018aa} in which the distance between RC1 and RC2 was estimated from their relative $K_s$ difference.

We also tested whether the difference in $\mu_l^*$ between RC1 and RC2 varies with longitude, as expected from differential Galactic rotation. To do this we constructed $\mu_l^*$ difference profiles by binning the data in longitude and comparing them with simulations. The observed profile broadly matches the theoretical trend, with differences becoming more negative towards positive longitudes.

We also found that the mean extinction difference between the two structures is approximately 5\,\% of the total extinction towards the Galactic bar along the same line of sight (i.e. RC1). This suggests that, beyond the Galactic bar, there is little interstellar material along the line of sight before reaching the potential spiral arm.

Additionally, we analysed the extinction curve in the NIR towards each of the VVV tiles in the target region by computing the slope of the RC features. We derived an extinction curve $A_J/A_{K_s}=3.34\pm0.07$, which is consistent with previous estimates near the GC \citep[e.g.][]{Alonso-Garcia:2017aa,Nogueras-Lara:2020aa,Sanders:2022wa}. We did not observe significant variation in the extinction curve across different lines of sight (i.e. tiles) or along the line of sight (for the RC features, which are proposed to be $\sim5$\,kpc apart, as suggested by \citealt{Gonzalez:2018aa}), within the uncertainties.

\begin{acknowledgements}
  
This work is based on observations made with ESO Telescopes at the La Silla Paranal Observatory under programme ID 179.B-2002. SJ gratefully acknowledges the financial support received through the MPIA Summer Internship Programme 2022. FN-L gratefully acknowledges the sponsorship provided by the European Southern Observatory through a research fellowship, as well as support from grant PID2024-162148NA-I00, funded by MCIN/AEI/10.13039/501100011033 and the European Regional Development Fund (ERDF) “A way of making Europe,” and from the Severo Ochoa grant CEX2021-001131-S, funded by MCIN/AEI/10.13039/501100011033. KF acknowledges financial support from the European Research Council under the ERC Starting Grant "GalFlow" (grant 101116226).

\end{acknowledgements}

  \bibliographystyle{aa} 
  \bibliography{BibGC.bib} 
\begin{appendix} 

\section{Extinction curve calculation details}

\setlength{\tabcolsep}{3.5pt}
\def\arraystretch{1.1}

\begin{table}[ht!]

\caption{Slope and extinction curve for each tile and RC feature.}
\label{table_extinction}

\begin{tabular}{c c c c c}
\hline\hline
Tile & Slope RC1 & Slope RC2 & $A_J/A_{K_s}$ RC1 & $A_J/A_{K_s}$ RC2 \\
\hline
b313 & $0.418 \pm 0.008$ & $0.410 \pm 0.011$ & $3.391 \pm 0.046$ & $3.437 \pm 0.063$ \\
b314 & $0.427 \pm 0.004$ & $0.423 \pm 0.006$ & $3.343 \pm 0.019$ & $3.361 \pm 0.035$ \\
b315 & $0.440 \pm 0.013$ & $0.438 \pm 0.010$ & $3.274 \pm 0.065$ & $3.281 \pm 0.051$ \\
b316 & $0.432 \pm 0.007$ & $0.436 \pm 0.011$ & $3.313 \pm 0.039$ & $3.294 \pm 0.058$ \\
b317 & $0.424 \pm 0.010$ & $0.424 \pm 0.018$ & $3.358 \pm 0.054$ & $3.360 \pm 0.100$ \\
b318 & $0.417 \pm 0.003$ & $0.419 \pm 0.003$ & $3.395 \pm 0.018$ & $3.387 \pm 0.020$ \\
b319 & $0.418 \pm 0.002$ & $0.417 \pm 0.003$ & $3.391 \pm 0.014$ & $3.401 \pm 0.015$ \\
b320 & $0.432 \pm 0.005$ & $0.437 \pm 0.011$ & $3.314 \pm 0.029$ & $3.286 \pm 0.056$ \\
b321 & $0.433 \pm 0.005$ & $0.430 \pm 0.006$ & $3.307 \pm 0.026$ & $3.325 \pm 0.035$ \\
b322 & $0.434 \pm 0.008$ & $0.433 \pm 0.002$ & $3.303 \pm 0.044$ & $3.309 \pm 0.013$ \\
b323 & $0.426 \pm 0.010$ & $0.425 \pm 0.012$ & $3.348 \pm 0.058$ & $3.352 \pm 0.068$ \\
b341 & $0.423 \pm 0.009$ & $0.420 \pm 0.011$ & $3.367 \pm 0.049$ & $3.380 \pm 0.063$ \\
b342 & $0.431 \pm 0.004$ & $0.431 \pm 0.006$ & $3.319 \pm 0.023$ & $3.318 \pm 0.033$ \\
b343 & $0.431 \pm 0.005$ & $0.434 \pm 0.006$ & $3.322 \pm 0.025$ & $3.302 \pm 0.030$ \\
b344 & $0.425 \pm 0.005$ & $0.423 \pm 0.005$ & $3.352 \pm 0.027$ & $3.364 \pm 0.030$ \\
b345 & $0.435 \pm 0.004$ & $0.431 \pm 0.010$ & $3.298 \pm 0.023$ & $3.318 \pm 0.054$ \\
b346 & $0.434 \pm 0.004$ & $0.433 \pm 0.005$ & $3.306 \pm 0.021$ & $3.312 \pm 0.027$ \\
b347 & $0.395 \pm 0.015$ & $0.395 \pm 0.010$ & $3.530 \pm 0.095$ & $3.530 \pm 0.062$ \\
b348 & $0.426 \pm 0.008$ & $0.421 \pm 0.005$ & $3.348 \pm 0.042$ & $3.373 \pm 0.027$ \\
b349 & $0.444 \pm 0.009$ & $0.442 \pm 0.010$ & $3.251 \pm 0.047$ & $3.264 \pm 0.053$ \\
b350 & $0.440 \pm 0.014$ & $0.428 \pm 0.018$ & $3.273 \pm 0.074$ & $3.335 \pm 0.096$ \\
b351 & $0.460 \pm 0.009$ & $0.456 \pm 0.020$ & $3.176 \pm 0.044$ & $3.193 \pm 0.094$ \\
\hline
\end{tabular}

\end{table}   

\end{appendix}

\end{document}